# Azimuthal rotation induced by the Marangoni force makes small Leidenfrost droplets move in random zig-zag directions

Ken Yamamoto[1,2]
山本 憲

[1] Department of Earth and Space Science, Osaka University, 1-1 Machikaneyama, Toyonaka, Osaka 560-0043, Japan
[2] Water Frontier Research Center (WaTUS), Tokyo University of Science, 6-3-1 Niijuku, Katsushika-ku, Tokyo 125-8585, Japan

**Abstract**
We observed the zig-zag motion of small Leidenfrost water droplets (radii less than 0.6 mm) on a hot, flat substrate. To understand this motion, we conducted an experiment using a glass capillary to fix a droplet at its edge and control the droplet height. Thermographic and interferometric observations reveal that the droplets rotated both vertically and azimuthally. Based on the characteristic frequency of the azimuthal rotation depending on the substrate temperature and droplet height, we developed a semi-empirical model considering unsteady Marangoni convection and its relaxation. We confirmed that our model can predict the characteristic time of the zig-zag motion of unfixed Leidenfrost droplets.

## I. INTRODUCTION

Under specific conditions, a liquid droplet can levitate on a substrate by the Leidenfrost effect [1]. When the temperature of the substrate is above a certain point, which is called the Leidenfrost temperature [2], the vapor generated from the droplet forms a thin layer beneath it, and the vapor pressure supports the droplet, preventing contact with the substrate. This distinct configuration exhibits the rich and complicated dynamics of Leidenfrost droplets, which have been intensively investigated over the past two decades [2–9], and the translational motion of these droplets has become a hot topic [10–12].

Because the Leidenfrost droplets do not contact the substrate, friction is negligible, and their motion is strongly affected by gravity. Nevertheless, Bouillant *et al.* [10] showed that it is not always that simple for small droplets: on a flat substrate with a carefully adjusted horizontality of ~0.1 mrad, water droplets with a radius of $R = 1.0$ mm translated in random directions, while droplets with a radius of $R = 2.0$ mm sensitively felt gravity and translated in a certain direction. Bouillant *et al.* also visualized flows inside droplets [10, 13] and found that the flow structure changed with the droplet size. This transition was explained by the droplet shape (aspect ratio), which was spherical for small droplets and flattened for large droplets owing to gravity. In large droplets ($R > 1.5$ mm), the flow structure is axisymmetric (containing pairs of countercurrents), whereas, in small droplets, symmetry breaking occurs, and the internal flow exhibits a rolling motion. Symmetry breaking has been extensively investigated by Yim *et al.* [11], both experimentally and numerically, and the results imply that the rolling motion is likely to be induced by the thermobuoyant (Rayleigh–Bénard) effect rather than the thermocapillary (Marangoni) effect. While the origin of symmetry breaking is still under discussion, the relevance of rolling motion to translational motion is more feasible. Bouillant *et al.* [10] experimentally confirmed that the direction of rolling motion corresponds to the translation direction; hence, a random direction of small droplets is the result of the randomness of the rolling direction. Moreover, they found that the base of the rolling droplet was slightly (a few milliradians) inclined toward the translation direction. This stable inclination was induced by the lubricating vapor flow beneath the droplet [12].

As discussed above, the direction of motion is either stochastic or deterministic, depending on the size of the droplet; however, moving droplets are considered to maintain their lines because of their frictionless nature. Therefore, to change or control the translational direction of Leidenfrost droplets, physical or thermal inclination [14] or non-uniformities of the substrate [15–18] were employed. However, in this study, we realized that submillimeter-scale water droplets ($R < 0.6$ mm) frequently change their translation direction on a flat substrate under uniform condition. To understand this new finding, we performed two different experiments and derived a semi-empirical model that characterizes this unique behavior.

## II. EXPERIMENTAL SETUP AND PROCEDURE

To understand the zig-zag motion of submillimeter droplets, we performed two different experiments: a submillimeter-droplet experiment (with droplets of $R < 0.6$ mm) and a pendant-droplet experiment (with millimeter-scale pendant droplets). The amount of contamination in millimeter-scale droplets was monitored throughout the experiment and we concluded that it was negligible. For submillimeter droplets, although the monitoring was difficult, we concluded that the contamination effect was negligible because the droplets did not exhibit sudden acceleration or "explosion" [19, 20], characteristics indicative of significant contamination.

### A. Submillimeter-droplet experiment

We observed the translational motion of submillimeter water droplets ($R < 0.6$ mm). The droplets were generated by atomization through the contact boiling of large droplets ($R \sim 2$ mm) on a flat horizontal silicon wafer (Fig. 1a). The polished silicon wafer (2 in.) was heated at 250°C on a glass heater (SG-101, BLAST). The horizontal orientation of the hot substrate was carefully adjusted using a level guide with a precision of < 1 mrad. The motion of the small Leidenfrost droplets was captured from a top-down view using a high-speed camera (CP-70-1-M-1000, Optronis) at 1000 fps *via* a macro lens (25 mm F2.8 2.5-5X ULTRA MACRO, LAOWA).

Large droplets ($R \sim 2$ mm) were dropped through a syringe from *ca.* 1 cm above the hot substrate to initiate contact boiling, resulting in nucleate boiling and generation of submillimeter droplets. Among tens of atomized droplets, we tracked a few of them that remained after most of the droplets disappeared. The selection criteria for the droplets included (i) the substrate temperature uniformity, (ii) independence of the droplet, and (iii) low acceleration. For criterion (i), we estimated the time to recover the substrate temperature from the contact boiling as $\tau_{\text{recover}} \sim R_L^2 \alpha^{-1} \sim 0.05$ s, where $R_L$ and $\alpha$ denote the radius of the large droplet and the heat diffusivity of the substrate, respectively. To assure the temperature uniformity, we took an interval before tracking $t_{\text{start}}$ of at least 0.5 s after the end of contact boiling. Criteria (ii) and (iii) were determined *a posteriori*. If the droplet motion was unaffected by adjacent droplets during the tracking, it was considered independent. If the droplet motion exhibited no significant acceleration or deceleration, the effects of the spread from the parent droplet was regarded as negligible. In addition to these criteria, the effect of the convective flows of vapor and air was estimated and regarded as negligible because the tracked droplets on the substrate did not exhibit collective motion (see Supplementary Movie 1).

The trajectories and sizes of the selected droplets were tracked using a lab-made Python program. The submillimeter droplets had non-zero translational velocity at the start of the tracking $t_{\text{track}} = 0$ and its typical velocity was $10^0$ mm s$^{-1}$. We evaluated the turning of the droplet by estimating the instantaneous turn rate $d\theta_D/dt$ by measuring the change in the angle of the moving direction at a certain time. Then, we obtained a characteristic period to turn 360° as $P_D = 2\pi (d\theta_D/dt)^{-1}$ for a droplet size averaged over the measurement time.

**B. Pendant-droplet experiment**

To achieve prolonged observation, droplets attached to the edge of a glass capillary were used (Fig. 1b). A hot circular flat sapphire substrate (thickness: 1 mm, diameter: 28.5 mm) was placed on a ring heater. The surface temperature of the substrate $T_s$ was maintained at a constant value using a PID temperature controller (SBX-303, Sakaguchi E.H Voc Corp.). We set $T_s$ = 220–320°C. Above the hot substrate, the glass capillary (outer diameter: 2.0 mm, inner diameter: 1.0 mm, length: 100 mm, the edge is manually polished) was held vertically, and its bottom end had a distance to the substrate $H$ of 1.0–2.5 mm, which was adjusted by a z-axis translation stage. The top end of the capillary was connected to a syringe pump (Pump 11 Elite, Harvard Apparatus) *via* Tygon tubing to infuse DI water (at room temperature) to the bottom end of the capillary, where a spherical segment droplet was formed. After starting the infusion, the water reached the lower end of the capillary and the three-phase contact line advanced on the surface of the capillary end until it was pinned at the capillary edge. The droplet then grew toward the hot substrate. The final shape of the droplet was determined using a combination of $T_s$, $H$, and the infusion flow rate $Q$. Once the droplet reached its final shape, it was maintained there for several minutes. Because the flow rate (which is roughly equivalent to the evaporation rate) was very low (~10 μL min$^{-1}$), the supplied water was sufficiently warmed in the capillary by heat conduction.

Using the aforementioned setup, the droplet surface temperature distributions were measured using an IR camera (PI640, Optris) at 100 fps. For selected cases, interferometry was performed to obtain interferometric images of the droplet base [6, 10, 21, 22]. For interferometry, a high-speed camera (AX-100, Photron), a 4× objective lens (CFI Plan Flour 4X, Nikon), an LED light source (INTENSILIGHT C-HGFI, Nikon), and a filter block [containing two band-pass filters (XF1410/25 and XF3401/25, Omega) and a half mirror (XF121/25.7*36, Omega)] were mounted on an inverted microscope (Ti-2, Nikon). Interference fringes were obtained by irradiating a blue beam extracted from the LED light source *via* the band-pass filter. The droplet surface temperature was measured using an IR camera (PI160, Optris) at 120 fps in synchronization with the interferometric images recorded at 500 or 1000 fps.

**III. RESULTS AND DISCUSSION**

**A. Droplet motion**

Submillimeter water droplets ($R$ < 0.6 mm) were generated by atomization. As shown in Fig. 2 and Supplementary Movie 1, these droplets spontaneously changed their direction of translation. Figure 2 shows that the straight-running stability tended to decrease with the droplet size, but the direction of the turn and the angle in which the turn was completed were random (for instance, they occasionally turned 270° to the left and then turned 90° to the right). To understand this new finding, we performed a pendant-droplet experiment in which the time-resolved droplet surface temperature and droplet-base shape were obtained.

Figure 3 shows results of the simultaneous measurement of the temperature distribution of the droplet surface and the interference fringes of the droplet base ($H$ = 1.0 mm, $T_s$ = 250°C). The surface temperature measurement (Fig. 3a) shows that the droplet has a hotspot on its side and rotates in the azimuthal direction (in the horizontal plane) with a certain characteristic period $P$. Moreover, we found that the hotspot always rotated in synchronization with the rotation of the interference fringes, which always showed line symmetry, although the rotational direction (clockwise or counterclockwise) appeared to be random and sometimes switched (Fig. 3b and Supplementary Movie 2). However, these rotational motions were not always observed.

While the rotation direction was random and the rotation occasionally switched, started, or ceased, the synchronization of the hotspot and fringes was always consistent. To characterize the rotation, we measured a time evolution of angle $\theta$, the angle of the axis of symmetry to the horizontal lines of the interference images, as shown in Fig. 3b. As shown in Fig. 3c, the change in $\theta$ is almost constant in time and we can derive a characteristic period of rotation $P$ = 0.98 s. Because an unfixed droplet moves in a specific direction corresponding to the direction of the symmetry axis of its base shape [10], we interpreted that this (perfect or imperfect) rotation of the droplet base shape caused a zig-zag motion of the droplet. In fact, the base profile inclined along the axis of symmetry (Fig. 3d) as observed by Bouillant *et al.* [10]. Furthermore, we observed that the direction of the inner flow, which was visualized by a few particles, corresponded to the axis of symmetry and rotated with the axis (Supplementary Movie 3).

As the above results suggest that the azimuthal rotations of the hot spot and droplet base were synchronized and that the rotation was key to understanding the zig-zag motion of the droplet, we measured the hotspot rotation with a higher spatial resolution (with the IR camera, PI640, Optris) at 100 fps. A clear vertical high-temperature belt, which was considered to stem from the Rayleigh–Bénard instability of the internal flow, was observed. Moreover, we measured the azimuthal rotation period $P$ of the hot belt for different droplet heights $H$ and substrate temperatures $T_s$ by measuring the frequency of the temperature increase at the center of the droplet. We found that $P$ was positively correlated with both $H$ and $T_s$ for 230°C $\leq T_s \leq$ 280°C (Fig. 4). The Leidenfrost state became unstable (contact with the substrate occurred) at $T_s$ = 220°C. We regarded a rotation of at least three successive rounds as a stable rotation. Although experiments at each condition were repeated at least three times, the stable rotation was observed only once for $T_s$ = 280°C, $H$ = 2.0 mm. Therefore, the results for this case were omitted hereafter. We also found that the stable rotation was suppressed for $T_s$ > 280°C and/or $H$ > 2.0 mm, although incomplete rotations (one, two, or fewer rotations) were observed in some of these cases.

**B. Temperature profile of the pendant droplets**

We measured the temperature line profiles of the droplets along the $z$-coordinate, whose direction was normal to the substrate, and the origin was located at the surface of the hot substrate (see the inset of Fig. 5a). Figure 5a shows the time evolution of the local temperatures, the $z$ positions of which are indicated by different colors. A periodic temperature rise owing to the azimuthal rotation of the hot belt was clearly observed. Moreover, it was found that the temperature was always a decreasing function of $z$ even in the hot belt (except close to the capillary edge). This implies that the droplet surface was constantly cooled by its ambient temperature and the hot belt was formed by a surface flow (or flow near the surface) from the bottom to the apex.

Because the above results suggest that the droplet surface always had a temperature gradient, we obtained time-averaged line profiles for $T_s$ = 230, 250, and 280°C in Figs. 5b, 5c, and 5d, respectively. As shown in Fig. 5, the measured temperature at a small $z$ is affected by radiation from the substrate, which is reflected at the droplet surface. Therefore, we applied line fitting to the temperature profile at $z \geq 1.0$ mm and the resulting temperature gradients $C_1$ were ~7 °C/mm at $T_s$ = 230°C, ~5 °C/mm at $T_s$ = 250°C, and ~5 °C/mm at $T_s$ = 280°C. These results suggest that the temperature gradient was almost insensitive to $T_s$, whereas the absolute value of the local temperature depended on $T_s$. Note that the liquid temperature near $z = H$ corresponded to that measured by a K-type thermocouple within ±2°C. This implies that the linear temperature gradient $C_1$ was inaccurate because it underestimated the temperature at $z = 0$ when extrapolated from the temperature at $z = H$ and the gradient $C_1$. This underestimation could be due to the internal flow. As shown in Fig. 5a, the temperature gradient varied significantly during the rotation period, and its maximum value was almost double that of the averaged gradient $C_1$.

**C. Development of a semi-empirical model**

In this section, we develop a physical model that explains the azimuthal rotation and the dependence of its period on $H$ and $T_s$, as shown in Fig. 4. Because the droplet base was inclined (Fig. 3d), we assumed that a vertically rotating flow was generated by the thermobuoyant effect. In addition to the thermobuoyant effect, a thermocapillary effect (generation of Marangoni force) should exist because of the temperature gradient on the droplet surface. The direction of the net Marangoni force from the bottom of the droplet should then be the direction in which the temperature gradient is at its maximum, perpendicular to the direction of the internal flow (hot belt), as shown in Fig. 6a. Furthermore, because the Marangoni force generates a surface flow (characterized by velocity $u_{Ma}$) from the bottom of the droplet to the cold region, the hot liquid is transported in a characteristic time, and the flow relaxes. The characteristic time of relaxation $\tau_{Ma}$, which is called the Marangoni relaxation time [23], is calculated as

$$\tau_{\text{Ma}} = \left[\frac{\rho_L L^3}{\left(\frac{\partial \gamma}{\partial T}\right) \Delta T}\right]^{1/2}, \quad (1)$$

where $\rho_L$, $L$, $(\partial \gamma / \partial T)$, and $\Delta T$ denote the liquid density, characteristic length, surface tension gradient against temperature, and the temperature difference between the starting and ending points of the flow, respectively.

Because the inner flow and the Marangoni flow are mutually perpendicular at the droplet base, a net velocity is calculated by a vector sum and its angle to the inner flow $\varphi$ (in the horizontal plane) is $\varphi \sim \arctan(u_{\text{Ma}}/u_{\text{in}})$ where $u_{\text{Ma}}$ and $u_{\text{in}}$ are the Marangoni flow velocity characterized as $\sim L\,\tau_{\text{Ma}}^{-1}$ and the velocity of the thermobuoyant flow, respectively. After a lapse in the relaxation time, a new cold region appears at an angle perpendicular to the direction of the net velocity in the previous period, and the Marangoni flow is regenerated. The regeneration of the Marangoni flow repeats and the hot belt exhibits the azimuthal rotation with a characteristic period $\Pi \sim 2\pi \arctan^{-1}(u_{\text{Ma}}/u_{\text{in}})\,\tau_{\text{Ma}}$.

We calculate $\tau_{\text{Ma}}$ by substituting $(T_w - T_{\text{sat}})$ for $\Delta T$ and $H$ for $L$, where $T_w$ denotes a liquid temperature at a specific region, where the droplet surface is normal to the IR camera, and $T_{\text{sat}}$ denotes the liquid saturation temperature, respectively. The result is shown in Fig. 6b in relationship to the azimuthal rotation period $P$. In Fig. 6b, $P$ has a linear relationship with $\tau_{\text{Ma}}$, which implies the proposed concept is plausible. Furthermore, we generalize $\tau_{\text{Ma}}$ by substituting the temperature gradient $C_1$. Choosing $C_1 = 5$ °C/mm, we can write $\Delta T = C_1 H$ and the Marangoni relaxation time is rewritten as

$$\tau_{\text{Ma}}^* \sim KH, \quad (2)$$

where

$$K = \left[\frac{\rho_L}{C_1\left(\frac{\partial \gamma}{\partial T}\right)}\right]^{1/2} \sim 32\ [\text{s m}^{-1}]. \quad (3)$$

As expected, $\tau_{\text{Ma}}^*$ shows a linear relationship with $P$ (Fig. 6c). Therefore, the characteristic period can be rewritten as follows:

$$\Pi^* \sim \frac{2\pi}{\arctan(K^{-1}/u_{in})} KH. \quad (4)$$

Eq. (4) suggests that $u_{\text{in}}$ and $H$ affect the rotational period. Because $P$ has already been shown to have a linear relationship with $H$, $u_{\text{in}}$ was investigated. Although it is evident from Fig. 6c that $u_{\text{in}}$ is a function of $T_s$, we also have to notice that the droplet shape changed with $H$ in our pendant-droplet experiment. Because of this setup, we first verified that $u_{\text{in}}$ [or, more precisely, $\arctan(K^{-1}/u_{\text{in}})$] was not affected by $H$ and then investigated the effect of $T_s$ on $u_{\text{in}}$. With regard to $u_{\text{in}}$, we only measured once during the interferometric measurement when few tracer particles were contained in a droplet. The measured $u_{\text{in}}$ was ~50 mm s$^{-1}$, which is a typical internal-flow velocity of Leidenfrost droplets [10] (See Supplementary Movie 3).

We estimated the effect of $H$ and $T_s$ on $u_{\text{in}}$ from the relationship between $H$ or $T_s$ and $P / \tau_{\text{Ma}}^*$, because $P / \tau_{\text{Ma}}^*$ is considered to be a sole function of $u_{\text{in}}$. Figure 7a shows the relationship between $P / \tau_{\text{Ma}}^*$ and $H / H_0$, where $H_0$ denotes the maximum height at which the rotation was observed ($H_0 = 2.0$ mm). The diagram indicates that $P / \tau_{\text{Ma}}^*$ is almost insensitive to $H$ and sensitive to $T_s$. This implies that the vertically rotating flow owing to the thermobuoyant effect was unchanged despite the shape change of the droplet. This result corresponds to Bouillant et al. [10], in which symmetry breaking (vertically rotating flow) occurred when the aspect ratio of the droplet (in their experiment, large droplets were deformed owing to gravity) changed from a value of order two to a value of order one. In our case, the highest aspect ratio was approximately 2 for $H = 1.0$ mm, which was still within the symmetry breaking range. Figure 7b shows the $T_s$ effect on $P / \tau_{\text{Ma}}^*$, with a normalized form of $(T_s - T_{\text{sat}}) / T_{\text{sat}}$ for the abscissa. This indicates that $u_{\text{in}}$ is proportional to $C_2[(T_s - T_{\text{sat}}) / T_{\text{sat}}]^n$. We drew a fitting curve using the least-square method and obtained the numerical constant $C_2 = 8.64 \pm 0.69$ and index $n = 2.74 \pm 0.16$. Using the fitting results, we can rewrite Eq. (4) as

$$\Pi^* \sim 8.64 \left[\frac{(T_s - T_{\text{sat}})}{T_{\text{sat}}}\right]^{2.74} KH. \quad (5)$$

The term $C_2[(T_s - T_{\text{sat}}) / T_{\text{sat}}]^n$ can include multiple effects, such as the $T_s$ effect on $u_{\text{in}}$ and $u_{\text{Ma}}$. Although the origin of this $T_s$ effect is unknown, it could be explained by an increase in the vapor flow rate, which would result in an increase in $u_{in}$, or by excess heat used for heating the liquid rather than for the phase change (see Supplementary Information). The possibility that the excess heat heated the liquid is implied in Fig. 5, where the temperature gradient tends to decrease as $T_s$ increases. A similar result was reported by Yim *et al.* [11], in which the gradient was a few degrees per millimeter at $T_s = 350°C$. As the Marangoni force is proportional to the temperature gradient, this tendency could work to increase $\Pi^*$ or even suppress the azimuthal rotation when the temperature gradient reaches zero.

We compared $P$ and $\Pi^*$ in Fig. 8a and found that the data collapsed on a 1:1 line. Finally, we measured the characteristic rotation period of the submillimeter droplets $P_D$ on the silicon wafer ($T_s = 250°C$) and compared it with the model. We defined $P_D$ as $2\pi$ divided by the change in the angle of the moving direction at a certain time. The measured $P_D$ against droplet diameter $D$ is plotted in the inset of Fig. 8b. Although the data were scattered, we were able to draw a fitting line of $P_D = C_3 [(T_s − T_{sat}) / T_{sat}]^{2.74} KD$ with a fixed $T_s = 250°C$ and $K = 32$. Consequently, we obtained the numerical factor $C_3 = 23.15 \pm 0.97$ and $P_D$ can be estimated by the following correlation (see Fig. 8b):

$$\Pi_D^* \sim 23.15 \left[\frac{(T_s - T_{sat})}{T_{sat}}\right]^{2.74} KD. \qquad (6)$$

Although the droplet shape of our designed experiment (attached to the capillary) and the freely moving submillimeter droplets (spherical) were different, the developed model can predict the characteristic azimuthal rotation period for both cases. The model also implies that a higher $T_s$ leads to a higher $u_{in}$ and that $u_{in}$ could be much higher than $u_{Ma}$ for a high $T_s$, which results in a longer rotation period. The droplet size is also important. For a relatively large $R$, $\Pi_D^*$ turns out to be long, and a large substrate will be required to observe the zig-zag motion. It is also remarkable that we barely observed circular motions (motions with a constant turn) of submillimeter spherical droplets on a flat plate, whereas steady azimuthal rotation was frequently observed for fixed droplets. This may be due to the difference in the boundary conditions—the local temperature of the capillary edge was affected by that of the neighboring liquid (Fig. S1). This effect may stabilize the rotation. This could explain the decreased stationarity of the rotation and the frequent changes in the rotational direction when $H$ became large, in which the temperature of the capillary edge remained almost unchanged. Frequent changes in direction were also observed when the capillary was replaced with a thin needle (see Supplementary Information and Supplementary Movie 4).

## IV. CONCLUDING REMARKS

In summary, we observed a zig-zag motion of submillimeter ($R < 0.6$ mm) water Leidenfrost droplets moving on a flat horizontal silicon wafer heated at 250°C. While the moving direction appeared to be random, their straight-running stability showed droplet-size dependence. To understand the mechanism of the zig-zag motion, we conducted an experiment in which the droplet was fixed at the edge of a glass capillary, and the effective droplet size could be controlled by varying the distance between the hot sapphire substrate and the capillary edge. We measured the droplet surface temperature using an IR camera and observed a hot vertical belt, suggesting the existence of a vertically rotating inner flow. Moreover, different substrate temperatures ($T_s = 220–320°C$) and the effective droplet sizes ($H = 1.0–2.5$ mm) revealed that the hot belt rotated in an azimuthal direction in a certain temperature range (230–280°C) and its characteristic period depended on $T_s$ and $H$.

A physical model based on Marangoni relaxation was proposed, and it successfully predicted the result after incorporating an empirical correlation that considered the $T_s$-dependence of the inner flow velocity $u_{in}$. Subsequently, characteristic azimuthal rotation period of the submillimeter (unfixed) droplets observed on the silicon wafer was measured. The proposed semi-empirical model predicted the motion of the unfixed droplets with a similar accuracy as that for the fixed droplets, despite the difference in their shapes (almost perfect sphere and part of the sphere). These findings suggest that the dynamics of small Leidenfrost droplets are asymmetric and are more complex than previously believed.

## ACKNOWLEDGMENTS

The author would like to thank Dr. Hiroaki Katsuragi (Osaka University), Dr. Yoshiyuki Tagawa (Tokyo

University of Agriculture and Technology), Dr. Koji Hasegawa (Kogakuin University), Dr. Yutaku Kita (King's College London), Dr. Hideaki Teshima (Kyushu University), and Dr. Kota Fujiwara (Central Research Institute of Electric Power Industry) for their fruitful discussions. The author thanks Dr. Masahiro Motosuke (Tokyo University of Science) for lending his experimental apparatus.

# FIGURES

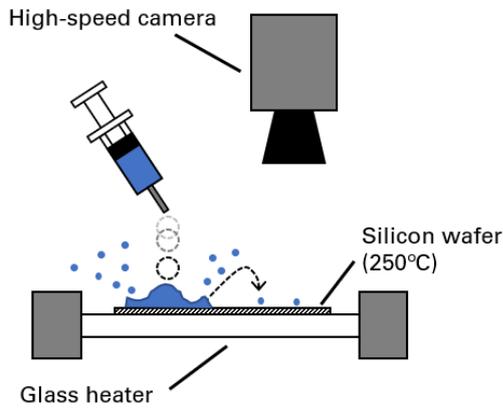
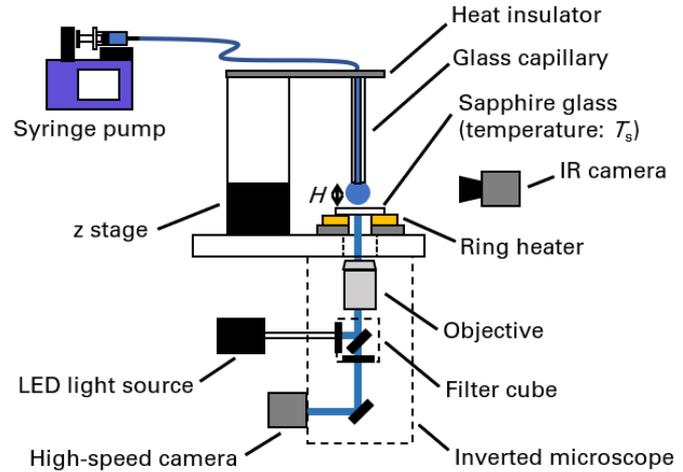

FIG. 1. Schematics of the experimental setup. (a) Submillimeter-droplet experiment. A polished flat silicon wafer is heated using a glass heater. A water droplet ($R \sim 2.0$ mm) is dropped onto the wafer to induce atomization through contact boiling and the translation of the atomized droplets are observed from the top-down view with a high-speed camera. (b) Pendant-droplet experiment. A pendant droplet is generated at one end of a glass capillary (2 mm in outer diameter, 100 mm in length) and the Leidenfrost effect is observed. The droplet height $H$ is changed by changing the distance from the capillary end to the hot substrate. The surface temperature of the droplet is measured using an IR camera. For selected cases, interferometric images are captured simultaneously.

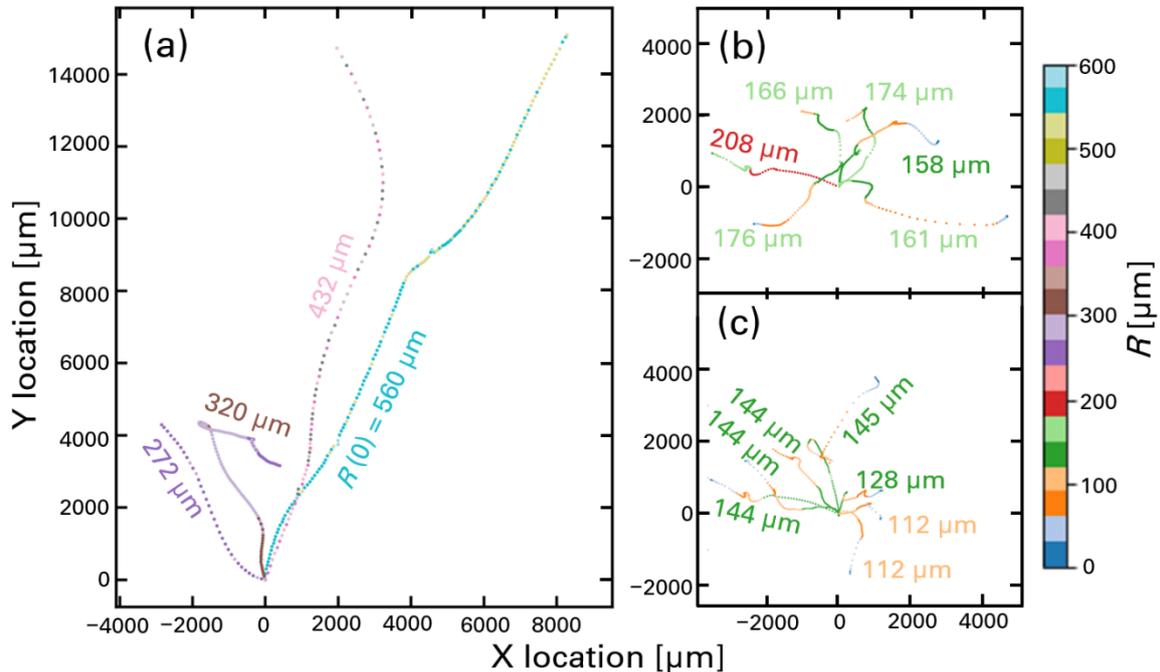

FIG. 2. Trajectory of small droplets translating on a flat horizontal silicon wafer ($T_s = 250°C$). The droplets are generated by atomization thorough the contact boiling of large droplets. The tracking is started at $t_{\text{track}} = 0$, at which time the target droplet is regarded as unaffected by other droplets and shows no significant acceleration or deceleration (but has a non-zero velocity). The droplet center is plotted with a 10-ms interval and the color bar indicates the radius of the droplet $R(t_{\text{track}})$. $(X, Y) = (0, 0)$ indicates the position at $t_{\text{track}} = 0$. $R(0)$ for each droplet is shown in the figure. Selected trajectories of (a) $R(0) = 560$–$272$ μm, (b) $R(0) = 208$–$158$ μm, and (c) $R(0) = 145$–$112$ μm, are shown. Initial translation directions are taken arbitrary for better visibility.

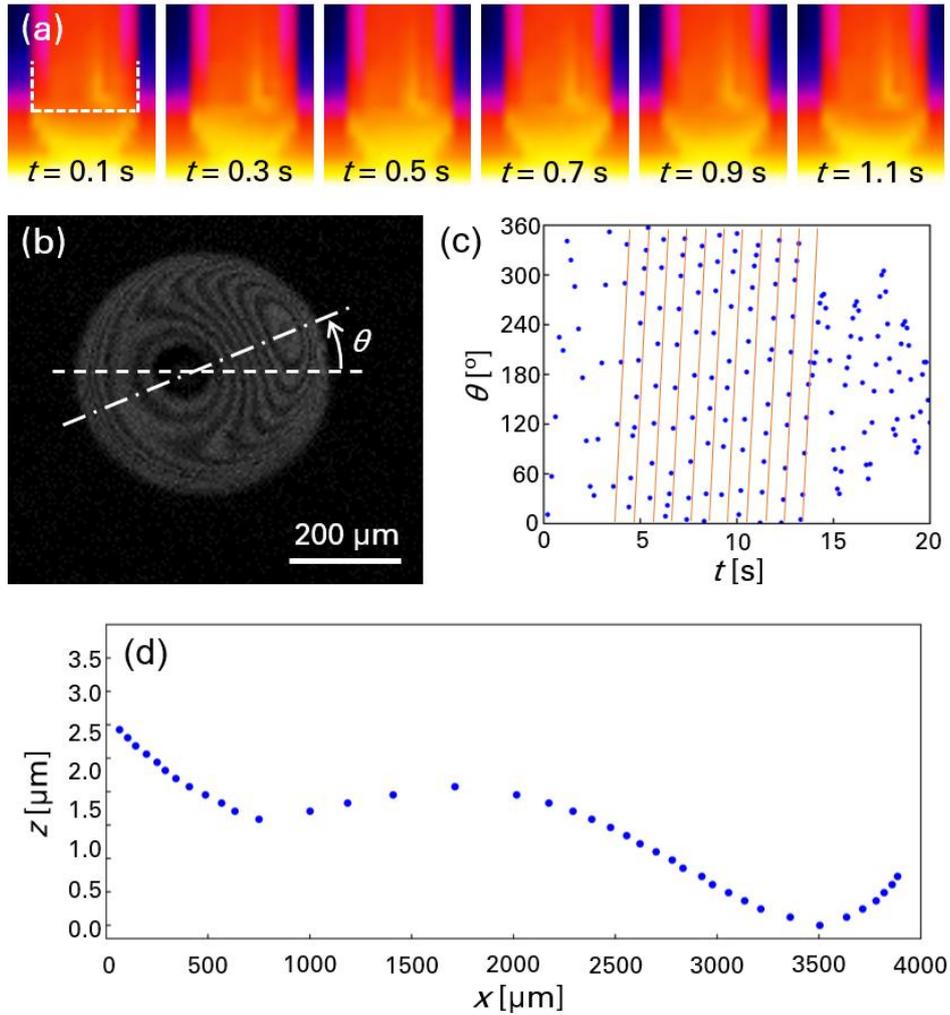

FIG. 3. Observation result of a droplet at $H = 1.0$ mm, $T_s = 250°C$. (a) Successive thermographic images. White dashed lines indicate the outline of the glass capillary (2.0 mm outer diameter). Image contrast is enhanced for better visibility. (b) Interference fringes of the droplet base (binarized after background subtraction). Angle of an axis of the line symmetry $\theta$ is defined. (c) Time evolution of $\theta$. Orange lines are guides showing the same inclination. The distance between the lines is 0.98 s. (d) Droplet-base profile along the axis of symmetry.

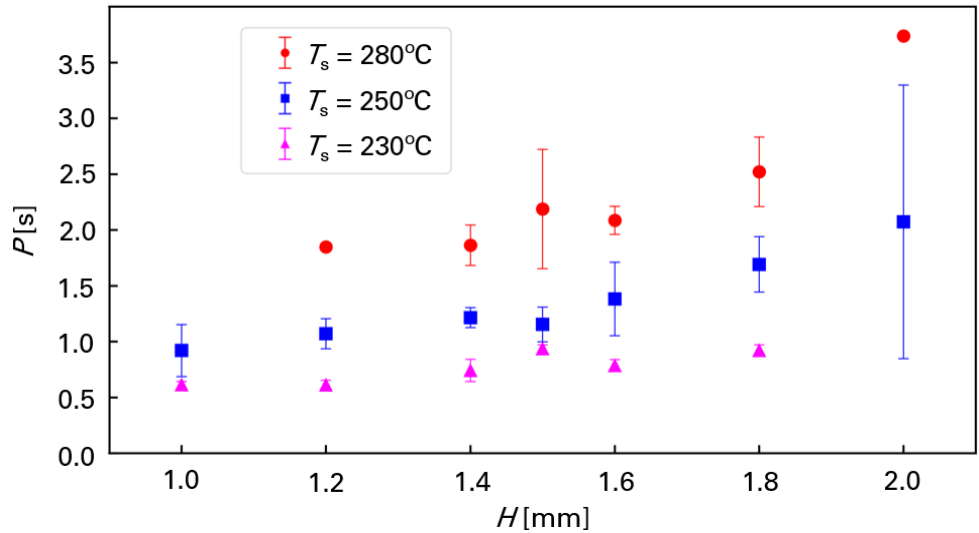

FIG. 4. Dependence of the characteristic rotation period $P$ on $H$ at three substrate temperatures $T_s$.

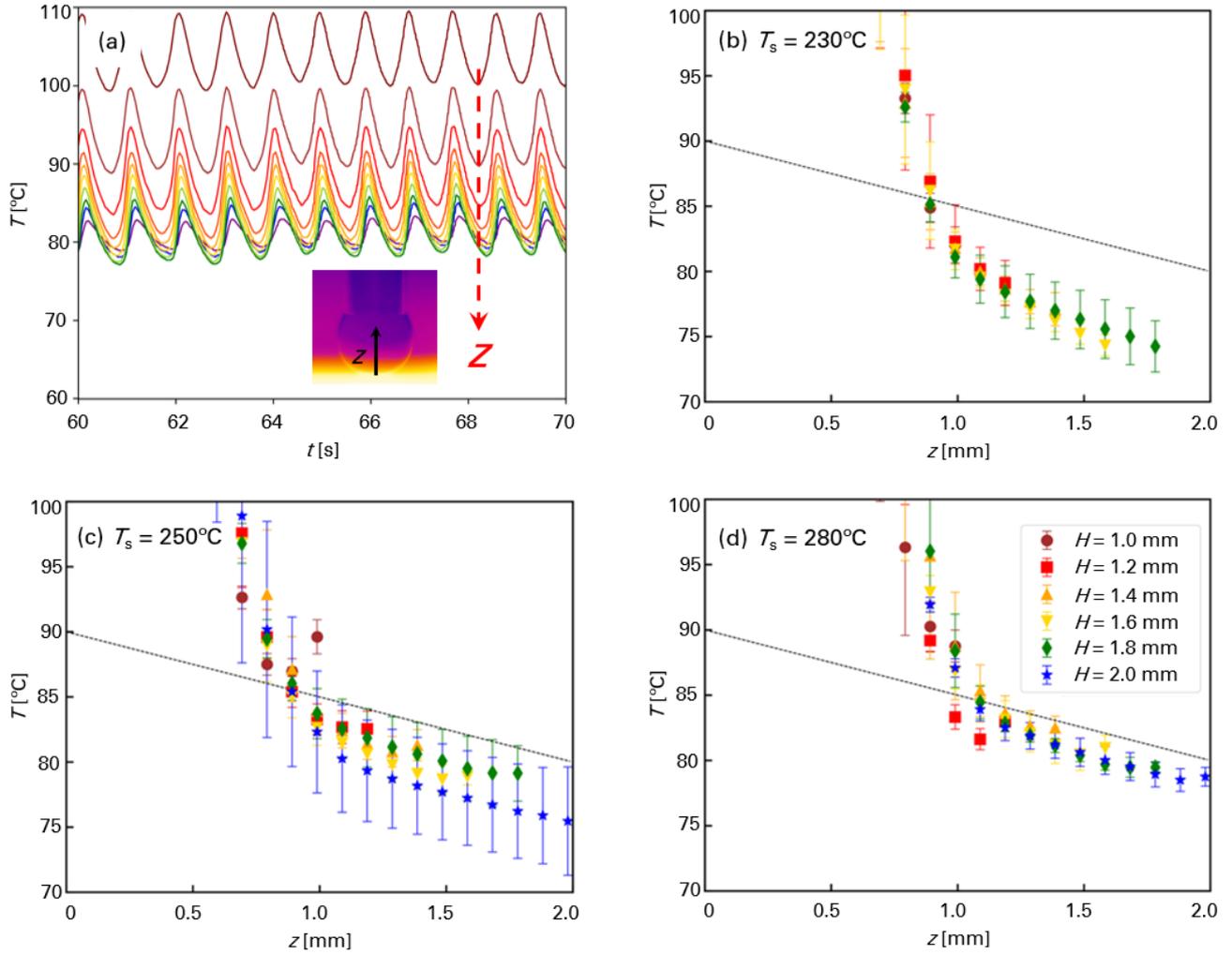

FIG. 5. (a) Time evolution of local temperature along a z-axis at $T_s$ = 250°C, $H$ = 2.0 mm (z-coordinate and its origin are indicated in the IR image). Relative $z$ position of the temperature curves is indicated in reverse order of the wavelength of color (purple: $z \sim H$). The $z$ distance between the curves is ~99 µm. (b)–(d) Time- and case-averaged temperature profiles along the z-axis for different $H$ and $T_s$. (b) $T_s$ = 230°C, (c) $T_s$ = 250°C, and (d) $T_s$ = 280°C. Black dashed lines indicate a slope of 5 °C/mm.

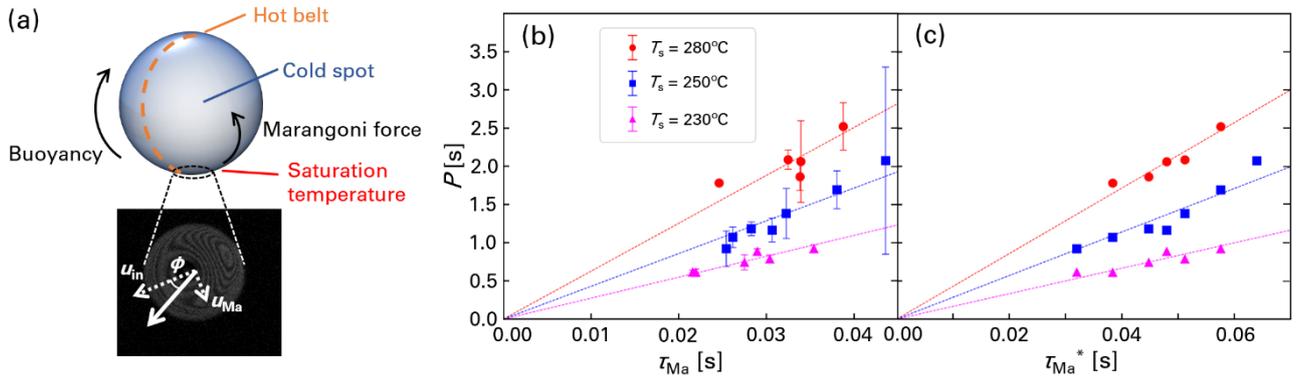

FIG. 6. (a) Schematic of the forces exerted on the droplet. Two perpendicular velocity vectors $u_{in}$ (stems from the thermobuoyant effect) and $u_{Ma}$ (stems from the thermocapillary effect) cause the azimuthal rotation of the hot belt. (b) $P$ vs. $\tau_{Ma}$. Dashed lines indicate linear fitting passing through the origin. (c) $P$ vs. $\tau_{Ma}^*$ (= $KH$). Dashed lines indicate linear fitting passing through the origin.

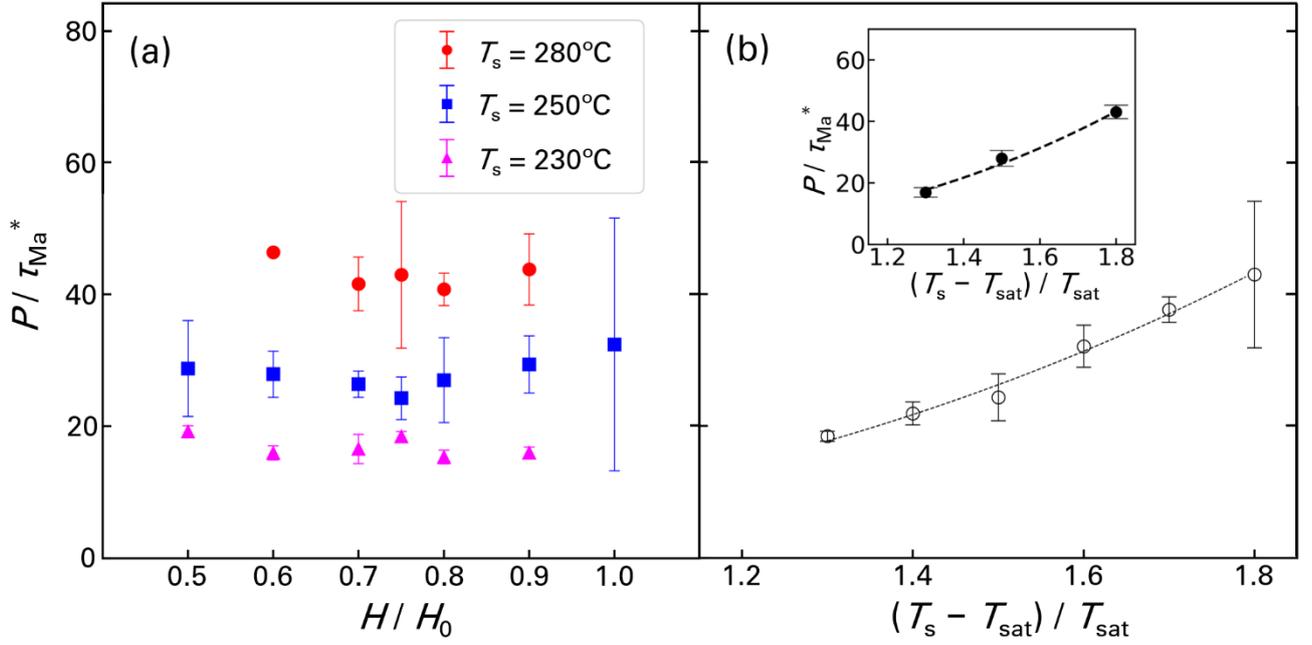

FIG. 7. (a) Relationship between $P$ normalized by $\tau_{Ma}^*$ and $H$ normalized by $H_0$ ($H_0$ = 2.0 mm). (b) $P / \tau_{Ma}^*$ vs. $(T_s - T_{sat}) / T_{sat}$ for $H$ = 1.5 mm. The dashed line indicates a fitting curve of $P / \tau_{Ma}^* = 8.64 \, [(T_s - T_{sat}) / T_{sat}]^{2.74}$. Inset is $P / \tau_{Ma}^*$ vs. $(T_s - T_{sat}) / T_{sat}$ for different $H$ at $T_s$ = 230, 250, and 280°C. Each plot is an average value of $P / \tau_{Ma}^*$ at $H$ = 1.0–2.0 mm. The dashed line indicates the same fitting curve.

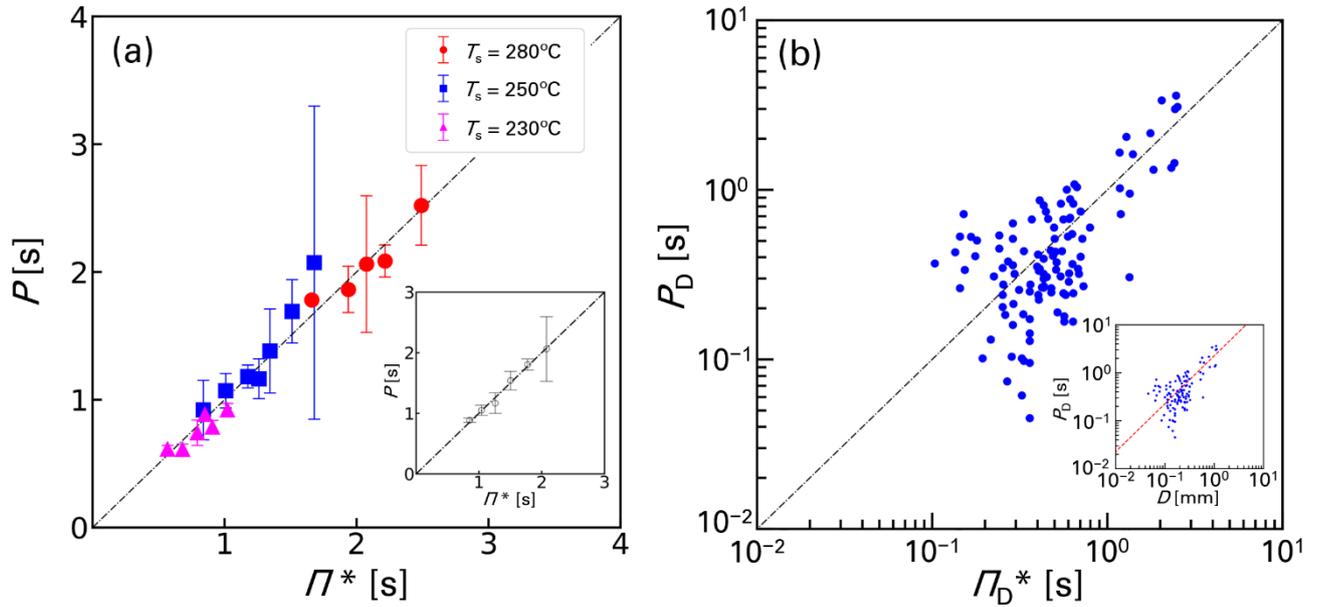

FIG. 8. Comparison of the predicted time with the measured characteristic time. Black dashed lines indicate 1:1 line. (a) Result of the droplets fixed at the capillary. The inset is the temperature dependence ($T_s$ = 230, 240, 250, 260, 270, and 280°C) at $H$ = 1.5 mm. (b) Result of the submillimeter droplets at $T_s$ = 250°C (log-log plot). The inset is the droplet-diameter dependence of $P_D$. The red dashed line indicates a fitting curve of $P_D = 23.15 \, [(T_s - T_{sat}) / T_{sat}]^{2.74} KD$. The coefficient of determination is 0.702.